# Video OCR for Video Indexing

Sankirti S. and P. M. Kamade

*Abstract*—Video OCR is a technique that can greatly help to locate the topics of interest in video via the automatic extraction and reading of captions and annotations. Text in video can provide key indexing information. Recognizing such text for search application is critical. Major difficult problem for character recognition for videos is degraded and deformated characters, low resolution characters or very complex background. To tackle the problem preprocessing on text image plays vital role. Most of the OCR engines are working on the binary image so to find a better binarization procedure for image to get a desired result is important. Accurate binarization process minimizes the error rate of video OCR.

## I. MOTIVATION

Video OCR is a technique that can greatly help to locate topics of interest in a large digital video archive via the automatic extraction and reading of captions and annotations [9]. Captions generally provide vital search information about the video being presented - the names of people and places or description of objects. Understanding the content of videos requires the intelligent combination of many technologies: speech recognition, natural language processing, search strategies, image understanding, etc. Extracting and reading captions provides additional information for video understanding. Performing Video OCR on video and combining its results with other video understanding techniques will improve the overall understanding of the video content. Although there is a great need for integrated character recognition in text-based video libraries. Automatic character segmentation was performed for titles and credits in motion picture videos in however; papers have insufficient consideration of character recognition. There are similar research fields which concern character recognition of videos. In character extraction from the car license plate using video images is presented and characters in scene images are segmented and recognized based adaptive thresholding. While these results are related, character recognition for the video presents its own difficulties because of different conditions of title character size and complex backgrounds. In video caption resolution of character is lower; also, the background complexity is more severe than in other research. The first problem is low resolution of the characters. The size of an image is limited by title number of scan lines defined in the NTSC standard; a character of the video caption are small to avoid occlusion of interesting objects such as people's faces. Therefore, the resolution of characters in the video caption is insufficient to implement stable and robust Video OCR systems. Another problem is the existence of complex backgrounds. Characters superimposed on videos often have hue and brightness similar to the background, making extraction extremely difficult. These problems in video OCR have opened an area for research.

Video OCR is a technique that can greatly help to locate topics of interest in a large digital video via the automatic extraction and reading of captions and annotations. Video OCR process and all the process modules required in video OCR are explained in section III. Applications of video OCR are explained in section IV. Conclusion based on relative work is explained in chapter V.

## II. PROBLEM STATEMENT

Performing Video OCR on video and combining its results with other video analysis techniques will improve the overall understanding of the video content. Recognition of videotext is a challenging problem due to various factors such as the presence of rich, dynamic backgrounds, low resolution, color, etc. A strategy is required to process the video images to produce high-resolution binarized text images that resemble printed text and minimize the error rate while recognition of degraded character.

## III. VIDEO OCR

In this section we review the maturity of the different component technologies that constitute an end-to-end video text recognition system. The purpose of this review is to identify thrust areas for research, and capabilities that may be ready for integration into a production environment. Figure 3.1 contains a block diagram that shows the processing sequence in a typical video text processing system. As with a traditional document processing system, the first step is to detect the text, i.e., find text in the image and demarcate the position of the text region.

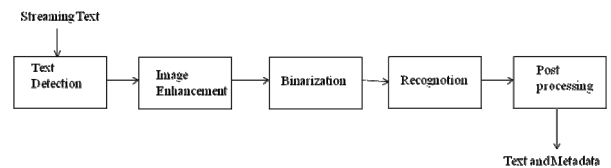

Figure 3.1 Steps in video OCR

The primary challenge being that we need to first track the sequence of instances of a video text object over successive frames of video, even in the presence of factors such as irregular, spasmodic motion of a handheld or vehicle mounted camera, nonlinearity of text plane motion, occlusions, shadows and other illumination changes, and so on. So, following the initial detection step, the next step is to enhancement. In addition, as part of the image enhancement step, video text images that exhibit perspective distortion (typically of text that is within the scene and not directly facing the camera); need to be rectified so that their appearance is similar to text that directly faces the camera. Text rectification involves the estimation of the position and orientation of the text plane relative to the camera. As most of the OCR's are working on the binarized image, binarization of enhanced image is required. Finally, binarized text is recognized by standard OCR and post-processing is applied on recognized text.

The task of detecting, segmenting and recognizing text visually appearing in complex images and/or video seems to be well defined. However, many design decisions have to be taken based on the overall goal.

Typical design choices are:
- What kind of text occurrences should be considered?

Based on its origin there exist two different kinds of text in videos and images. Scene text is text that was recorded as part of scene such as street names, shop names, and text on T-shirts. It mostly appears accidentally and is seldom intended. In contrast; the appearance of overlay text is carefully directed. It is often an important carrier of information and herewith suitable for indexing and retrieval.
- With what font attributes?

Text occurrences can differ significantly in font size, type, style, and color. Some research work has been tailored to very specific domains with limited variations in these attributes.
- In what kind of media data?

Should the underlying text detection, segmentation and recognition approach be image-based (i.e., treating a video as a set of independent images) or should it exploit the fact that the same text line occurs in videos for some time and that, therefore, the multiple instances of the same text line can be utilized to achieve better detection, segmentation and recognition performance.
- How will the output of the Video OCR system be used?

Different usages have different levels of tolerance against errors. For instance, if the Video OCR output is only used for image/video indexing based on the transcribed text, pixel errors in the localization and segmentation steps as well as recognition errors can be tolerated and compensated. If, however, the output is used for object-based video encoding, the system must minimize the errors in pixel classification. The system in, for example, was explicitly designed to label each pixel in a video as whether it belongs to text or not.

*A. Text Detection*

Most of the existing video text detection methods have been proposed on the basis of color, edge, and texture based features [2].

1. Color-based approach

Color-based approaches assume that the video text is composed of a uniform color. In the approach by the red color component is used to obtain high contrast edges between text and background. In (Hua et al, 2004) the "uniform color" blocks within the high contrast video frames are selected to correctly extract text regions.

2. Edge-based approach

Edge-based approaches are also considered useful for overlay text detection since text regions contain rich edge information. The commonly adopted method is to apply an edge detector to the video frame and then identify regions with high edge density and strength

3. Texture-based approach

Texture-based approaches, such as the salient point detection and the wavelet transform, have also been used to detect the text regions.

*B. Enhancing the Text Image*

Image Enhancement is used to improve the overall quality of an image, so that the result is more suitable than the original image for specific application. A typical characteristic of text in video is that a given text region persists over a few frames of video feed during which the background may or may not vary. In fact, more often than not the background varies while the text remains static. The enhanced image is computed by aligning the different instances of a particular text region across frames and, for each pixel, choosing the color that corresponds to the minimum intensity value across frames. We can try other order statistics such as the mean, median, and the maximum but the minimum order statistic yielded the best image in terms of visual perception.

*C. Binarization*

Most OCR engines are working on binarized text image for recognition. There are various different methodologies available for it. Finding the better one among them is important to get accurate result is important. Objective of this report is to study the available methods and combine them to get a required result

*Correlation-Based Technique*

A correlation technique for binarizing videotext images is one of the most popular conventional techniques [3]. Four separate filters to model horizontal strokes, vertical strokes and two diagonal strokes are used to find various strokes. The filters are trained by marking suitable regions on sample training data. Examples of marked training regions for horizontal and vertical filters and corresponding trained filters are shown below in figure 3.2 [5]. Each videotext image is correlated separately with each of the four filters, and the correlation outputs are thresholded to yield four intermediate binary images. The final binarized image is the union of the four intermediate binarized images.

Problem with this systems are –
i. The most prevalent problem is that background components occasionally exhibit text-stroke-like characteristics and are routinely picked up by the filters. Furthermore, such background components have high correlation scores and cannot be completely eliminated by adjusting the threshold.
ii. Constraining the binarization procedure to use the grayscale image alone, valuable information in the color image is summarily discarded.
iii. The binarized images also lack the smoothness that characterizes the curves and loops in characters such as **c, d, 0,** etc.

Notwithstanding all of the problems listed above, the correlation technique does an excellent job of locating the position of text pixels. Thus, while the final binarized image may be morphologically lacking, it does contain most of the text pixels in the original image [4]. Based on this observation a new binarization scheme can be developed that uses the correlation method as the first step and then reverts back to the color image for improved performance.

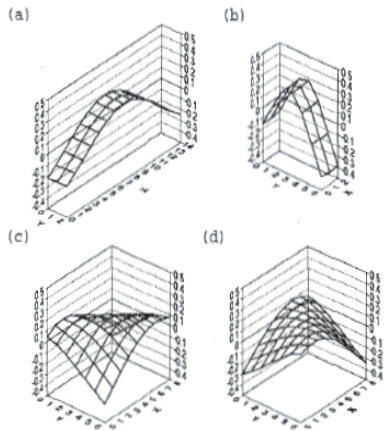

Figure 3.2 Character extraction filters (a) Vertical. (b) Horizontal. (c) Left diagonal. (d) Right diagonal [8].

*D. Recognition*

Most OCR engines ultimately work on binarized images of text. In other words, even if the input image is grayscale or color, the recognition system first converts the image into a bi-level (black-white) image before processing the image through the recognition engine.

*E. Post-processing*

To acquire text information for content-based access of video databases, high word recognition rates for video OCR are required. Once the text for a frame has been recognized, it is stored to be compared to the text extracted from neighboring frames for indexing. We apply post-processing, which evaluates differences between recognition results with words in the dictionary, and selects a word having the least differences. Different post-processing techniques are used for indexing. Video indexing can be used in various applications like digital libraries, digital News.

## IV. APPLICATIONS OF VIDEO OCR

Performing Video OCR on video and combining its results with other video understanding techniques will improve the overall understanding of the video content. Text in video provides rich information for content based search applications. There are various applications in which video OCR is used. Figure 4.1 shows application of video OCR where text in video is detected and meaning of that text can be derived which can be used in various applications.

1. Automatic broadcast annotation: creates a structured, searchable view of archives of the broadcast content.
2. Digital media asset management: archives digital media files for efficient media management.
3. Video editing and cataloguing: catalogs video databases on basis of content relevance.
4. Library digitizing: digitizes cover of journals, magazines and various videos using advanced image and video optical character recognition (OCR).
5. Mobile visual sign translation: extracts and translates visual signs or foreign languages for tourist usage, for example, a handhold translator that recognizes and translates Asia signs into English or French.
6. Named matching with face: Name and title information is valuable in matching the people's face with their name.

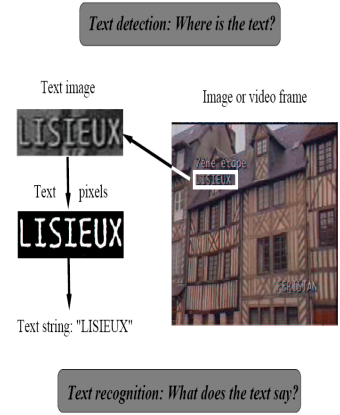

Figure 4.1 Application of video OCR.

## V. CONCLUSIONS

Most OCR engines, including ultimately work on binarized images of text. In other words, even if the input image is grayscale or color, the recognition system first converts the image into a bi-level (black-white) image before processing the image through the recognition engine. In the case of video text images, the attributes listed earlier (e.g., low resolution, perspective distortions, compression artifacts) make the binarization step a challenging one. In fact, experimental results indicate that a single binarization approach may not be adequate for dealing with different kinds of text in video, and a hybrid technique that combines multiple approaches offers most promise.